\documentclass[%
 reprint,
superscriptaddress,
 amsmath,amssymb,
 aps,
twocolumn]{revtex4-2}

\usepackage{graphicx}
\usepackage{dcolumn}
\usepackage{bm}
\usepackage[dvipsnames]{xcolor}
\usepackage{stackrel}
\usepackage{siunitx}

\usepackage{amsfonts}
\usepackage[english]{babel}
\makeatletter
\@namedef{l@en}{\l@english}
\@namedef{captionsen}{\captionsenglish}
\@namedef{daten}{\dateenglish}
\makeatother
\usepackage{bbold}

\newcommand{\DN}{\Delta N}
\newcommand{\brak}[1]{\langle #1 \rangle}
\newcommand{\Nmean}{\brak{N}}

\usepackage{hyperref}
\usepackage[shortlabels]{enumitem}

\usepackage{multirow}

\usepackage{lineno}
\newcommand{\tadv}{t_{\rm adv}}
\newcommand{\tdiff}{t_{\rm diff}}
\newcommand{\DNtt}{\langle \Delta N^2(t) \rangle}
\newcommand{\vaoup}{\boldsymbol{v}_{\rm OU}}
\newcommand{\dotvaoup}{\dot{\boldsymbol{v}}_{\rm OU}}

\AtBeginDocument{%
  \setlength{\linenumbersep}{\dimexpr\oddsidemargin+0.48in-\linenumberwidth\relax}%
}

\begin{document}

\preprint{APS/123-QED}

\title{Number fluctuations distinguish different self-propelling dynamics}

\author{Tristan Cerdin}%
\affiliation{Université Paris-Saclay, CNRS, FAST, 91405, Orsay, France}
\affiliation{
CNRS, Sorbonne Université, Physicochimie des Electrolytes et Nanosystèmes Interfaciaux, F-75005 Paris, France}

\author{Sophie Marbach}%
\email{sophie.marbach@cnrs.fr}
\affiliation{
CNRS, Sorbonne Université, Physicochimie des Electrolytes et Nanosystèmes Interfaciaux, F-75005 Paris, France}

\author{Carine Douarche}
\affiliation{Université Paris-Saclay, CNRS, FAST, 91405, Orsay, France}

\begin{abstract}
In nonequilibrium suspensions, static number fluctuations $N$ in virtual observation boxes reveal remarkable structural properties, but the dynamic potential of $N(t)$ signals remains unexplored. Here, we develop a theory to learn the dynamical parameters of self-propelled particle models from $N(t)$ statistics. Unlike traditional trajectory analysis, $N(t)$ statistics distinguish between models, by sensing subtle differences in reorientation dynamics that govern re-entrance events in boxes. This paves the way for quantifying advanced dynamic features in dense nonequilibrium suspensions.
\end{abstract}

\maketitle



The remarkable diversity of kinetics exhibited by living and synthetic self-propelled particles, at both individual and collective scales, continues to captivate physicists. These systems offer a unique window into the fundamental principles of active matter \cite{marchetti2013hydrodynamics, Ramaswamy2010, Bechinger2016}. For example, microorganisms like bacteria navigate their environment via dynamic swimming patterns, such as the well-known ``run and tumble'' motion of \textit{Escherichia coli}. These bacteria combine persistent motion with abrupt stochastic reorientations, a process that enables efficient exploration and adaptation \cite{Berg1983, Dahlquist1972, Schnitzer1993, Saragosti2011}. This seemingly simple behavior gives rise to complex collective phenomena, from swarming to pattern formation \cite{Vicsek1995, Tailleur2008}. Synthetic particles, such as Janus colloids, achieve self-propulsion through local chemical reactions or external fields, often coupled with diffusive reorientation~\cite{howse_self-motile_2007, Buttinoni2013, Palacci2013, Ebbens2011}. The interplay between persistent motion and random reorientation in these systems mirrors, yet distinctively contrasts with, the strategies used by their biological counterparts\cite{Cates2015}. Understanding how these diverse kinetic modes translate into collective dynamics, and how they can be distinguished through observable signatures — such as number fluctuations — remains a central challenge \cite{Deseigne2010,fily2012athermal}. 

%

\begin{figure}[ht!]
    \centering
    \includegraphics[width=0.99\linewidth]{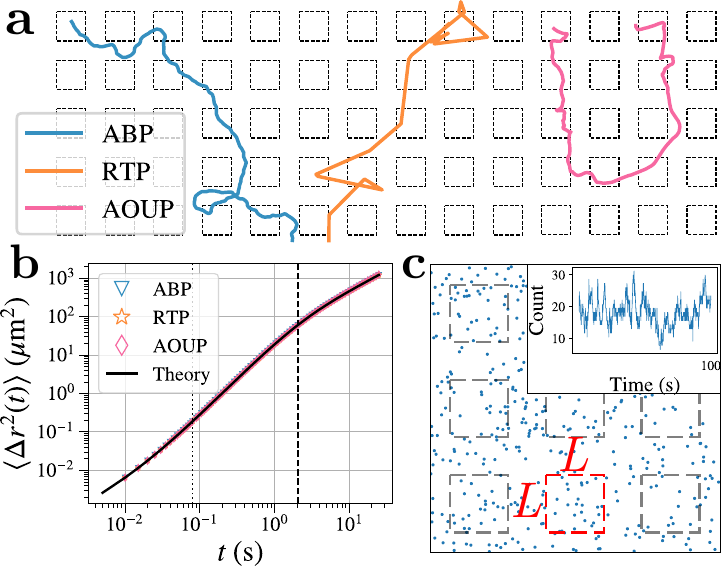}
    \caption{\textbf{Probing self-propelled particle models.}  (a) Simulated trajectories of 3 models -- Active Brownian (ABP), Run and tumble (RTP) or Active Ornstein-Uhlenbeck Particles (AOUP) -- overlayed with virtual observation boxes of size $L = 3~\unit{\mu m}$. Here $v = 3~\unit{\mu m.s^{-1}} $, $D_r = 1~\unit{s^{-1}}$, and for illustrative purposes we turn off translational diffusion. (b) The mean-squared displacements for many particles are indistinguishable between different models and captured by a single theory Eq.~\ref{eq:MSD}. Here $v = 5~\unit{\mu m.s^{-1}} $, $D_t = 0.1~\unit{\mu m^2.s^{-1}}$ and $D_r = 1~\unit{s^{-1}}$. (c) ``Countoscope'' approach where we probe the number of particles in virtual observation boxes of a simulation. }
    \label{fig:fig1}
\end{figure}

Particle trajectory analysis is a useful tool to characterize individual dynamics in dilute systems~\cite{manzo2015review,rose2020particle}. While its effectiveness can be challenged by trajectory reconstruction ambiguities in denser environments, it remains a foundational approach in the field; for instance, in helping to illustrate the different limiting regimes in time of self-propelled motion - see Figs.~\ref{fig:fig1}a and b. 
The study of intermediate scattering functions (ISFs), obtained \textit{e.g.} via image correlations in Fourier space (ISFs) offers an established alternative for microorganisms \cite{martinez2012differential,kurzthaler2024characterization,kurzthaler2018probing,germain2016differential,wilson2011differential}. Although ISFs can require deeper interpretation in Fourier space~\cite{kurzthaler_intermediate_2016,zhao_quantitative_2024}, they continue to offer valuable tools to quantify self-propelled motion. 
Building on these approaches, there is an opportunity to develop novel statistical observables that further enhance our ability to characterize and understand complex active systems. Here, we explore how particle number fluctuations can serve as a diagnostic tool, revealing the dynamics driving self-propelled systems and their emerging behavior.



Recently, particle number fluctuations $N(t)$ in virtual observation boxes -- ``Countoscope'' -- were used to characterize diffusion coefficients in dilute and dense colloidal suspensions at equilibrium~\cite{carter2025measuring,mackay2024countoscope} (Fig.~\ref{fig:fig1}c). This approach relies solely on particle positions, avoiding trajectory reconstruction. It also operates in real space, enabling direct links between quantitative measurements and observations. In spite of its introduction more than a century ago by Smoluchowski~\cite{smoluchowski1916studien,smoluchowski1914studien}, the potential of this approach for nonequilibrium systems remains unexplored. 

In this work, we demonstrate how number fluctuations $N(t)$ can detect nonequilibrium dynamic features of self-propelled particles, focusing on the dilute regime. Using an analytical theory detailed in our companion paper~\cite{cerdin2026countoscope}, we show that the correlation functions of $N(t)$ probe the three primary dynamic regimes: short-term diffusion, intermediate-term self-propulsion, and long-term enhanced diffusion~\cite{bailey2022fitting}. We establish quantitative scaling laws to extract key dynamical properties. More importantly, we demonstrate $N(t)$ correlations can distinguish between different reorientation behavior.

We consider 3 broadspread models of self-propelled particles to benchmark number fluctuations in this nonequilibrium context. First, the run and tumble particle (RTP) model, which describes well some bacteria or microalgae~\cite{berg_e_2004, kurzthaler2024characterization, clement_bacterial_2016, polin2009chlamydomonas}. In 2D, the position $\bm{r}(t)$ of a particle with orientation $\bm{u}(\theta) = (\cos{\theta}, \sin{\theta})$ evolves as
\begin{align}
    &\dot{\boldsymbol{r}}(t) = v\boldsymbol{u}(\theta_i) + \sqrt{2D_t} \boldsymbol{\eta}(t) \label{eq:roft}\\
    & \theta \underset{\alpha}{\rightarrow} \theta' \in [0,2\pi] \label{eq:theta}
\end{align}
where $v$ is the self-propulsion speed, $D_t$ a translational diffusion coefficient, and $\bm{\eta}(t)$ a Gaussian white noise satisfying $\langle \bm{\eta}(t)\rangle=0$ and $\langle \eta_i(t)\eta_j(t')\rangle= \delta_{ij}\delta(t-t')$ where $\langle \cdot \rangle$ represents an average over noise. Eq.~\eqref{eq:theta} corresponds to exponentially distributed tumbling events at rate $\alpha$. We assume tumbling events are instantaneous. 
Second, we consider the Active Brownian Particle (ABP) model, which describes better systems where reorientation is smooth and diffusive, such as Janus particles~\cite{kurzthaler2018probing, poncet_pair_2021,sprenger_active_2020}. The evolution of $\bm{r}(t)$ is the same as for RTP Eq.~\eqref{eq:roft}, but orientation satisfies
\begin{equation}
    \dot{\theta}(t) = \sqrt{2D_r} \eta_r(t)
\end{equation}
where $\eta_r(t)$ is another Gaussian white noise and $D_r$ is an angular diffusion coefficient. 
When the self-propulsion speed changes with time, its fluctuations are captured by the Active Ornstein Uhlenbeck model (AOUP), as
\begin{align}
    &\dot{\bm{r}}=\vaoup(t)+\sqrt{2D_t}\bm{\eta}(t) \\
    &\tau_r\dotvaoup= -\vaoup(t) + \sqrt{2D_v}\bm{\eta}_v(t)
\end{align}
where $\tau_r$ is a timescale associated with velocity relaxation, $D_v$ a velocity diffusion coefficient and $\bm{\eta}_v(t)$ another Gaussian random noise. The mean particle velocity is $v \equiv \sqrt{\langle |\vaoup^2(t)| \rangle} = \sqrt{2D_v/\tau_r}$, allowing us to build correspondence with the ABP and RTP models. 

Particle motion is usually quantified by the mean squared displacement (MSD) $\langle \Delta \bm{r}^2(t) \rangle  = \langle |\bm{r}(t)-\bm{r}(0)|^2 \rangle $. The MSD of all 3 models is captured by a single equation (black line in Fig.~\ref{fig:fig1}a):
\begin{equation}
    \langle \Delta \bm{r}^2(t) \rangle  = 4 \left(D_t +\frac{v^2}{2D_r}\right)t + 2\frac{v^2}{D_r^2}(e^{-D_rt} -1 )
    \label{eq:MSD}
\end{equation}
where $D_r = \alpha $ for RTPs and $D_r = 1/\tau_r$  for AOUPs. Eq.~\eqref{eq:MSD} captures 3 regimes of motion: at short times, motion is diffusive $\langle \Delta \bm{r}^2(t) \rangle = 4D_t t$ until a crossover time $\tadv = 4D_t/v^2$, after which motion is ballistic $\langle \Delta \bm{r}^2(t) \rangle  = v^2 t^2$. After enough reorientation events, quantified through another crossover time $\tdiff = 4 D_{\rm eff}/v^2$ motion is diffusive again $\langle \Delta \bm{r}^2(t) \rangle = 4D_{\rm eff}t$  with an enhanced diffusion coefficient  $D_{\rm eff} = D_t +\dfrac{v^2}{2D_r}$. In many practical cases $D_{\rm eff} \gg D_{\rm t}$ and thus $\tdiff \simeq 2 \tau_r$. Notably, other statistical properties such as the velocity autocorrelation function are also similar for all 3 models. 

\begin{figure}[ht!]
    \centering
    \includegraphics[width=0.9\linewidth]{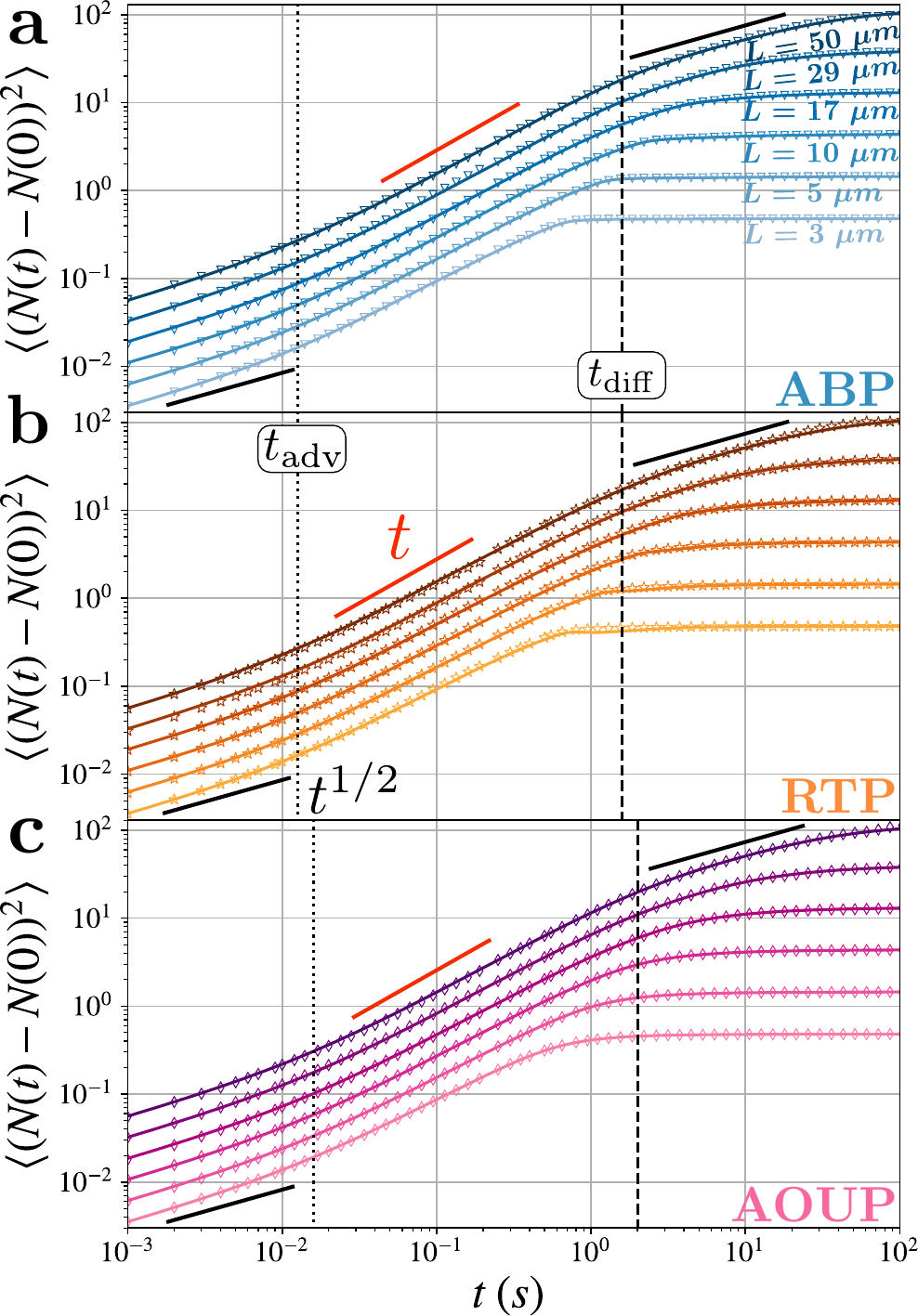}
    \caption{\textbf{Number fluctuations $\langle \Delta N^2(t) \rangle$ exhibit 3 distinct regimes in time,} for (a) Active Brownian particles, (b) Run and Tumble particles, and (c) Active Ornstein-Uhlenbeck particles. Box sizes are the same for all 3 plots, and go from small (in light color) to large boxes (in dark color). Parameters are the same as in Fig.~\eqref{fig:fig1}b. Symbols: simulation; lines: theory -- see text.}
    \label{fig:Countoscope_fig}
\end{figure}

We now investigate the signatures of self-propulsion on number fluctuations $N(t)$ (Fig.~\ref{fig:fig1}b) by simulating non-interacting suspensions of particles for all 3 models. We partition the simulation domain into finite square observation boxes of size $L$ (Fig.~\ref{fig:fig1}c) and count the particle number $N(t)$ in each box. We analyze the fluctuations $\langle \Delta N^2(t)\rangle$, defined as:
\begin{equation}
\begin{split}
\label{eq:nMSD}
    \langle \Delta N^2(t)\rangle  &= \brak{\left(N(t) - N(0)\right)^2} \\
    &= 2 (\brak{N^2} - \Nmean^2) -  2 C_N(t),
\end{split}
\end{equation}
where $C_N(t) = \brak{N(t) N(0)} - \Nmean^2$ is the correlation function for particle number, and the average $\langle \cdot \rangle$ is taken over boxes and initial times. Further details on spatial partitioning are in Ref.~\cite{carter2025measuring}. Results for each model and various box sizes are shown as symbols in Fig.~\ref{fig:Countoscope_fig}. Note that in this non-interacting case, at fixed time $t$, the particle number in a box is a Poisson process. Hence the number variance is proportional to the mean number $\langle N^2\rangle - \langle N \rangle^2 = \langle N \rangle$. 

\begin{figure}[ht!]
    \centering
    \includegraphics[width=0.99\linewidth]{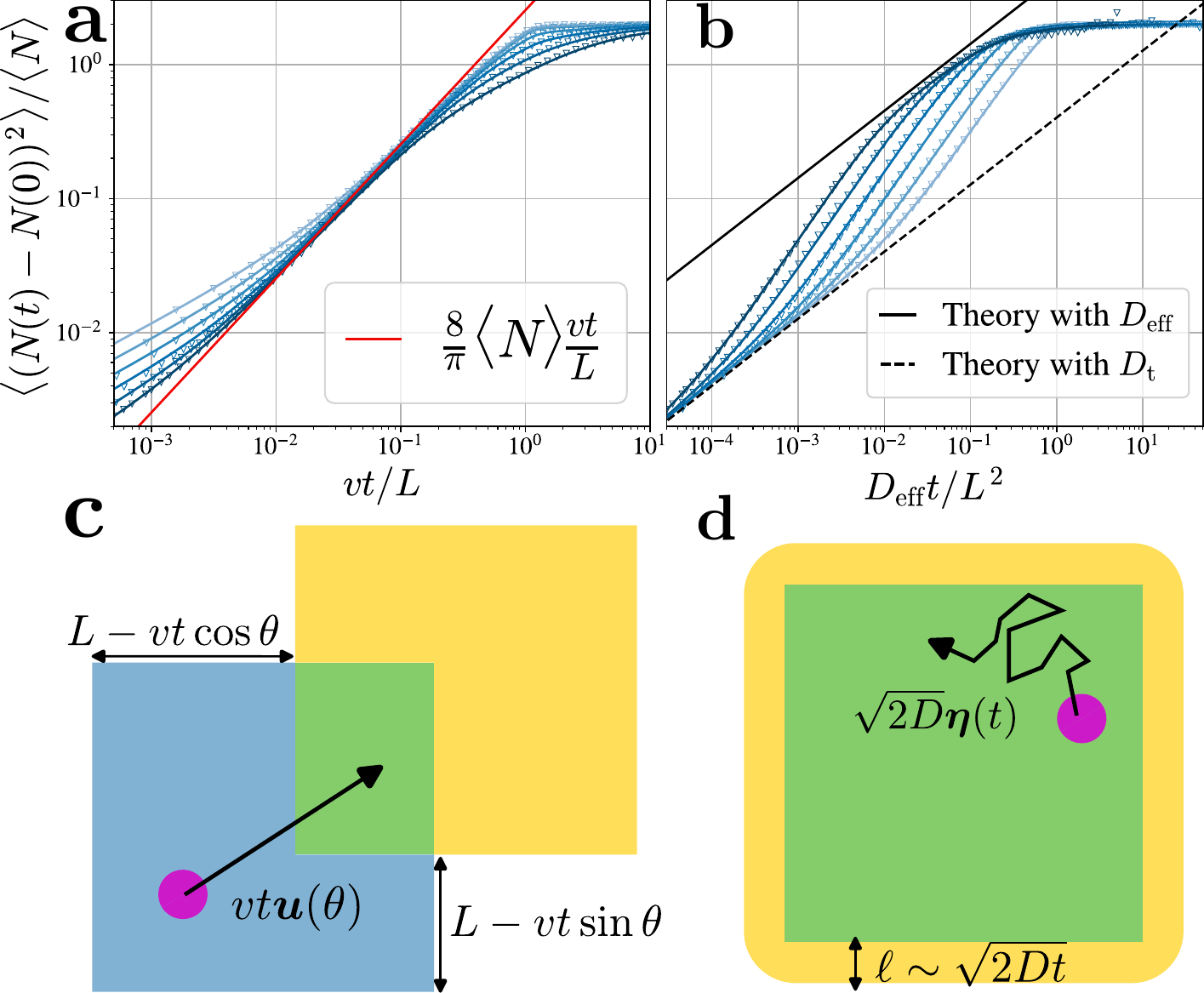}
    \caption{\textbf{Limit regimes of advection or diffusion captured by scaling laws.} Number fluctuations for ABPs, same data as in Fig.~\ref{fig:Countoscope_fig}a, with (a) advective rescaling in time and (b) diffusive rescaling. (c-d) Schematics illustrating the probability distribution clouds to find a particle in a box after some time $t$ (yellow) given it started in the box initially (blue) in a purely (c) advective or (d) diffusive case, with $D = D_t$ or $D_{\rm eff}$.}
    \label{fig:limiting regimes}
\end{figure}

The number fluctuations in Fig.~\ref{fig:Countoscope_fig} are clearly sensitive to all microscopic dynamics. 
Over time, fluctuations increase: as particles move away from their initial positions, it becomes increasingly more likely to sense number configurations different from the initial value. Eventually they plateau as the current state becomes uncorrelated from the initial state. Like MSDs, these fluctuations exhibit three distinct regimes, scaling as $\sqrt{t}$, $t$, and $\sqrt{t}$. The final $\sqrt{t}$ scaling is more apparent in large boxes, as for small boxes, fluctuations plateau faster and the final $\sqrt{t}$ regime does not have time to establish.
The diffusive and advective behaviors can be further highlighted by rescalings. The amplitude of fluctuations increases with larger boxes: it should scale like $\langle N\rangle$ in the absence of interactions. Rescaling time by $L/v$, the time to traverse a box with self-propulsion, curves collapse at intermediate times (Fig.~\ref{fig:limiting regimes}a). Rescaling time by the typical diffusion time, $L^2/D_{\rm eff}$, collapses curves at short and long times (Fig.~\ref{fig:limiting regimes}b).

To shed further light on these behaviors, we derive analytical theory for the 3 cases. Briefly, $N(t)$ is the integral over a box of the local particle density $\rho(\bm{r},t)$. Thus,
\begin{equation}
\label{eq:Cnt}
    C_N(t)= \frac{\langle N \rangle}{L^2}\iint_{[-\frac{L}{2},\frac{L}{2}]^2} d\boldsymbol{r}d\boldsymbol{r'} \int \frac{d\boldsymbol{k}}{(2\pi)^2}e^{i\boldsymbol{k}\cdot (\boldsymbol{r}-\boldsymbol{r'})} F(\boldsymbol{k},t)
\end{equation}
where $F(\boldsymbol{k},t) = \frac{1}{N}\langle \hat{\rho}(\boldsymbol{k},t) \hat{\rho}^*(\boldsymbol{k},0) \rangle$ is the density correlation function -- the ISF -- at wavenumber $\bm{k}$.
Using appropriate expressions of the ISFs, Eqs.~\eqref{eq:nMSD} and \eqref{eq:Cnt}, we obtain predictions for $\langle \Delta N^2(t)\rangle$. If particle motion is Gaussian, as is the case for AOUPs, then we simply have
\begin{equation}
    \langle \DN^2(t) \rangle = 2 \langle N \rangle \left( 1 - \left[ f\left( \frac{\langle \Delta \bm{r}^2(t) \rangle}{L^2}\right) \right]^2 \right) 
\label{eq:NMSD}
\end{equation}
where $\langle \Delta \bm{r}^2(t) \rangle$ is the MSD in Eq.~\eqref{eq:MSD} and $f(x \equiv \langle \Delta \bm{r}^2(t) \rangle/L^2) = \sqrt{x/\pi} \left( e^{-1/x} - 1\right) + \mathrm{erf} \left( \sqrt{1/x} \right)$. In the case of passive motion ($v = 0$), Eq.~\ref{eq:NMSD} recovers the results of Ref.~\cite{mackay2024countoscope}. For non Gaussian cases like ABPs and RTPs, exact albeit complex expressions can be found, and we report details in Ref.~\cite{cerdin2026countoscope}. These analytical expressions are presented as lines in Fig.~\ref{fig:Countoscope_fig} and are in perfect agreement with the simulations.  
Eq.~\eqref{eq:NMSD} illustrates that number fluctuations sense similar properties as the MSD. 

The different scaling laws in Fig.~\ref{fig:limiting regimes} call for physical insights.
To do so we follow Smoluchowski's original work~\cite{smoluchowski1916studien} and reexpress 
\begin{equation}
    \langle \DN^2(t) \rangle  = 2 \langle N \rangle  P_{\mathrm{out}}(t) \label{eq:probaout}
\end{equation}
where $P_{\mathrm{out}}(t)$ is the probability that a particle starting in the box at $t = 0$ is outside the box at time $t$. The factor 2 comes from the fact that it is as likely to start in the box and exit it than to start outside the box and enter it. Consider a particle anywhere with uniform probability in the box at $t = 0$ (blue square in Fig.~\ref{fig:limiting regimes}-c). If the particle only propels with speed $v$, at time $t$ it has moved away from the box in the propulsion direction $\theta$ (yellow square). The probability $\Psi_{\rm out} (\theta,t)$ that the particle is outside the box at $t$ is given by the ratio of the yellow area over the blue square, 
\begin{equation}
    \Psi_{\rm out} (\theta,t) =
        1 - \frac{(L-v|\cos{\theta}|t)(L-v|\sin{\theta}|t)}{L^2} \\
\end{equation}
where we supposed that $t$ is short enough that there is still a non zero probability for the particle to be in the box.
$P_{\mathrm{out}}(t)$ is then simply obtained by averaging over orientations $\theta$, as $P_{\mathrm{out}}(t) = \frac{1}{2\pi} \int_0^{2\pi} \, d\theta \,\Psi_{\rm out} (\theta,t)$. At short times $P_{\rm out} \simeq 4 v t/\pi L$ which gives
\begin{equation}
    \langle \Delta N^2(t)\rangle  \simeq \frac{8}{\pi} \langle N \rangle \frac{v t}{L}.
    \label{eq:intermediate}
\end{equation}
This scaling law agrees perfectly with our simulated data (red line in Fig.~\ref{fig:limiting regimes}a). Eq.~\eqref{eq:intermediate} holds only for ABP and RTPs, since for AOUPs we expect an average should also be done over the orientation and the amplitude of the velocity $\vaoup$. Since each velocity component of an AOUP follows a normal distribution, the amplitude $v$ of the 2D velocity follows a Rayleigh distribution, and we can obtain
\begin{equation}
    \langle \Delta N^2(t)\rangle  \simeq \frac{4}{\sqrt{\pi}} \langle N \rangle \frac{v t}{L}.
    \label{eq:intermediateAOUP}
\end{equation}
Alternatively, expanding Eq.~\eqref{eq:NMSD} at short times and taking $D_t = 0$ also recovers Eq.~\eqref{eq:intermediateAOUP}. This latter scaling law correctly reproduces intermediate behavior for AOUP (see Fig. S1).
The difference in prefactors between Eq.~\eqref{eq:intermediate} and Eq.~\eqref{eq:intermediateAOUP} already highlights how numbers sense differences between models.


\begin{figure*}[t!]
    \centering
    \includegraphics[width=0.99\linewidth]{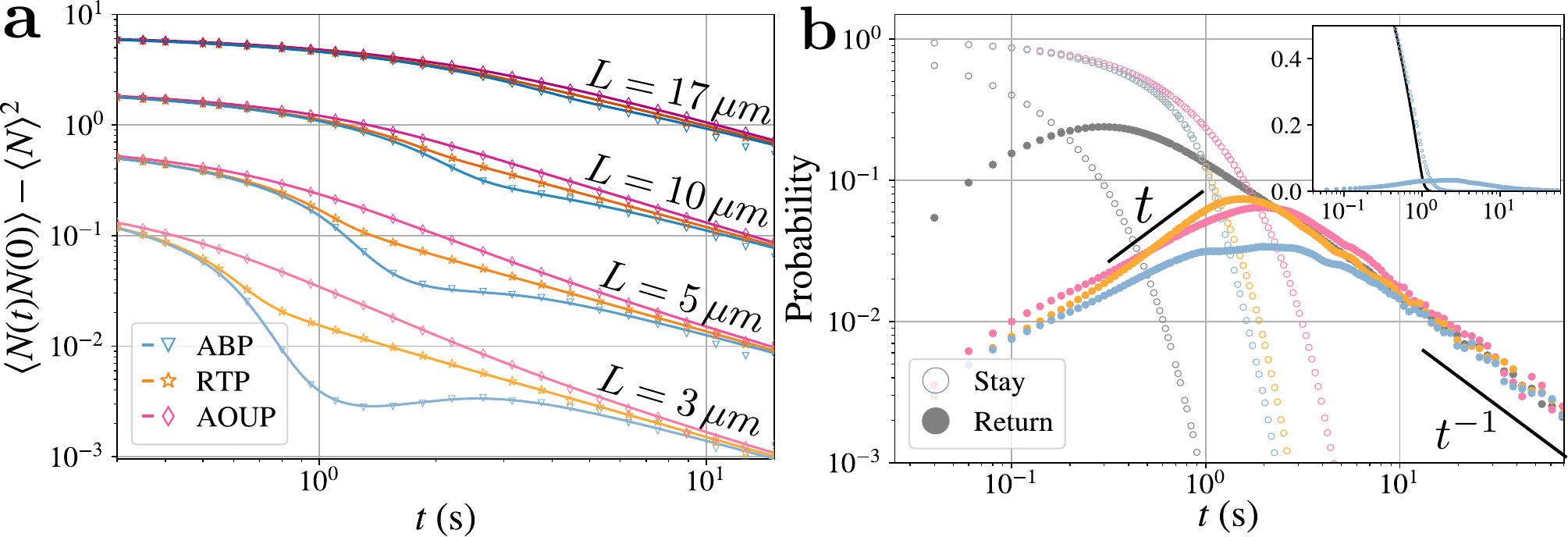}
    \caption{\textbf{Number correlation functions $C_N(t)$ distinguish different models.} (a) Number correlations from simulations (symbols) and theory (lines) for different box sizes; (b) $P_{\rm stay}(t)$ (open) and $P_{\rm return}(t)$ (full symbols) from simulations, taking $L = 5~\unit{\mu m}$. (inset) Lin-log scale of the same plot keeping just ABP data. Color codes for particle models are the same as in (a) and the gray color represents a passive case with diffusion coefficient $D_{\rm eff}$. Numerical parameters are the same as in Fig.~\ref{fig:fig1}b.}
    \label{fig:correlations}
\end{figure*}

Focusing now on diffusive behavior, take $v = 0$ and suppose the particle diffuses with diffusion coefficient $D$. We take $D$ a generic diffusion coefficient that could be $D_t$ or $D_{\rm eff}$ according to the typical timescales investigated. After some time $t$ the particle will have explored a thin region beyond the box of thickness $\ell \simeq \sqrt{2D t}$ (yellow band in Fig.~\ref{fig:limiting regimes}d). $P_{\rm out}(t) $ is then equal to the yellow area over the blue square area, $P_{\rm out}(t) \simeq 4 \ell L/L^2  = 4\sqrt{2Dt/L^2}$, yielding the appropriate $\langle \DN^2(t) \rangle \propto \sqrt{t}$ in the diffusive case. To obtain the correct prefactors, one has to expand Eq.~\eqref{eq:NMSD} at short times, assuming $\langle \Delta \bm{r}^2(t) \rangle = 4 D t$, and get
\begin{equation}
\langle \DN^2(t) \rangle \simeq \, \frac{4}{\sqrt{\pi}} \sqrt{\frac{4D t}{L^2}}.
    \label{eq:fluct1}
\end{equation}
Assuming $D = D_t$ captures the behavior of number fluctuations at short times Eq.~\eqref{eq:fluct1} (Fig.~\ref{fig:limiting regimes}b, dashed black line); and $D \rightarrow D_{\rm eff} = D_t + v^2/2D_r$ the one at long times (full black line). 

The transition times between translational diffusion and advective regimes $\tadv$, and then from advective to long time diffusive $\tdiff$ are thus slightly different between models. For the number fluctuations of RTPs and ABPs we find $\tadv = \pi D_t /v^2$ and $\tdiff = \pi D_{\rm eff}/v^2$, while $\tadv = 4 D_t/v^2$ and $\tdiff = 4 D_{\rm eff}/v^2$ for AOUPs. The discrepancies between crossover times for the different models yields a slight shift of number fluctuations behaviors (see Fig.~\ref{fig:Countoscope_fig}, vertical lines), towards later times for AOUPs. For RTPs and ABPs, the transition towards long time diffusive regime is sensed by $\langle \DN^2(t) \rangle$ about $25\%$ faster than in the MSDs, which could help resolve reorientational dynamics better~\cite{bailey2022fitting}.
Overall however, at the level of number fluctuations $\langle \DN^2(t) \rangle$, the 3 models are hard to distinguish (see Fig. S2).

However, a change of perspective can isolate differences between the 3 models. Instead of $\DNtt$, we focus on the autocorrelation function $C_N(t) = \langle N(t) N(0) \rangle - \langle N \rangle^2 = \langle N \rangle - \langle \DN^2(t) \rangle/2$. Although it is directly related to the number fluctuations via Eq.~\eqref{eq:nMSD}, it is more sensitive to long time behavior~\cite{carter2025measuring}, and thus potentially to reorientation dynamics. In Fig.~\ref{fig:correlations}a we present $C_N(t)$ for different box sizes and different models. Strikingly, sharp differences are visible between the 3 models, and predominantly so in small boxes. With our numerical parameters, $\tdiff\simeq 2~\mathrm{s}$ which is about the time where significant differences arise, and which is also comparable to the mean reorientation time $ \tau_r = 1~\mathrm{s}$. The correlation function for ABP exhibits a sharp dip, which corresponds to a temporary loss of correlations. This dip is not as apparent for RTP. For AOUPs correlation functions decay smoothly, without any local loss of correlation. 

To uncover why these differences emerge, we notice that $C_N(t)$ can be reexpressed as
\begin{equation}
    C_N(t) = \langle N \rangle (1 - P_{\rm out}(t)) = \langle N \rangle \left( P_{\rm stay}(t) + P_{\rm return}(t) \right)
\end{equation}
where $P_{\rm stay}(t)$ is the probability that the particle has remained in the box between $0$ and $t$, while $P_{\rm return}(t)$ is the probability that the particle escaped the box at least once between $0$ and $t$ and is back in the box at $t$. We calculate $P_{\rm stay}(t)$ and $P_{\rm return}(t)$ from our simulations and present them for a representative box size and all 3 models in Fig.~\ref{fig:correlations}b. We check that they indeed sum up to the correlation function in Fig. S3. Clearly, sharp differences are notices between all 3 models. 

To start, we interpret the shared qualitative features of  $P_{\rm stay}(t)$ and $P_{\rm return}(t)$.
For all models, at short times, $P_{\rm return}(t) \simeq 0$ as a particle has not escaped the box yet, and thus $P_{\rm stay}(t) \simeq 1 - P_{\rm out}(t)$. Supposing $P_{\rm out}(t) \simeq 4 v t/\pi L$ as for Eq.~\eqref{eq:intermediate}, we plot $1 - 4 v t/\pi L$ as a black line in the inset of Fig.~\ref{fig:correlations}. We obtain good qualitative agreement with numerical data for $P_{\rm stay}(t)$, such that for this small box, first exit events are dominated by advection out of the box. Around the reorientation time $\tau_r = 1~\mathrm{s}$, $P_{\rm return}(t)$ increases, reaching a maximum, before it eventually decays. At long times, $P_{\rm return}(t) \sim 1/t$ since, as any diffusive process in 2D, we expect return probabilities to decay as $1/t^{d/2}$ where $d = 2$ is the system's dimension~\cite{Voss1976,mackay2025collective}.

We now interpret differences. For ABP (blue) and RTP (orange), $P_{\rm stay}(t)$ decays similarly, while for AOUP (pink), $P_{\rm stay}(t)$ decays more slowly. This is due to fluctuations in an AOUP's velocity: an AOUP stays longer in a box before the first exit. This is quite apparent when looking at the example AOUP trajectory in pink in Fig.~\ref{fig:fig1}a. The most differences appear in $P_{\rm return}(t)$. The slowest particle to return to a box is the ABP, and the one that returns the most is the RTP. When an ABP particle leaves a box, it is quite ``slow'' to turn around, since it reorients via smooth diffusion. Such events are so rare that the blue ABP trajectory in Fig.~\ref{fig:fig1}a does not exhibit any returns. This explains the local loss in correlations. In contrast, an RTP will make sharp turns, allowing it to come back to a box more -- indeed we see about 4 returns in the orange RTP in Fig.~\ref{fig:fig1}a. Regardless of the model, such differences can only appear if the typical time to cross the box via advection, $L/v$ is smaller than the reorientation time $\tau_r$. The dips should thus only be visible when $L \leq v/D_r$. For our simulated data, $v/D_r = 5~\unit{\mu m}$ and indeed, dips are quite visible for box sizes $L \lesssim 5~\unit{\mu m}$. Number correlation functions are thus  sensitive to differences in reorientation dynamics.


Our investigation of number fluctuations echoes investigations on mean first passage times (MFPT) or survival probabilities, for which investigation is ongoing for self-propelling systems~\cite{iyaniwura2025mean,debnath2021escape,di2023active,mori2020universal}. In a disk with an absorbing boundary, the decay of the survival probability was observed to be mostly modulated by activity and not so much by reorientation dynamics~\cite{di2023active} -- which is similar to the behavior of $P_{\rm stay}(t)$ which characterizes part of the number fluctuations. Such differences in reorientation dynamics also play critical roles in triggering specific collective dynamics~\cite{rupprecht2016velocity}. While RTP returns more to a box here than ABP, in other contexts, this trend is reversed. For instance, in a confining potential, an RTP will more easily exit the confining potential than an ABP~\cite{solon2015active}. Number fluctuations thus provide an alternative way to shed light on the exotic properties of self-propelled systems. Further reorientation patterns such as run-reverse~\cite{detcheverry2017generalized} or random reorientations~\cite{gentili2025anomalous} should also exhibit specific signatures in the number correlations. 


The investigation of \textit{static} number fluctuations $N$ in observation volumes has proven useful in quantifying \textit{static} properties of dense non-equilibrium systems for over 30 years. For instance,
``giant'' number fluctuations, where $\alpha > 0$ in the scaling $\langle N ^2 \rangle - \langle N \rangle^2 \sim N^{1 + \alpha}$, indicate long-range organization in bacterial or synthetic
active matter suspensions~\cite{zhang2010collective,peruani2012collective,liu2021density,fily2012athermal,dey2012spatial,alarcon2017morphology,chate2008collective,chepizhko2021revisiting,fadda2023interplay,narayan2007long,toner2005hydrodynamics,navarro2015clustering}.  It is thus clear that investigating \textit{dynamic} number fluctuations in virtual observation boxes like $N(t)$, is a promising route to quantify collective dynamic features of self-propelling suspensions.

\vspace{2mm}
\acknowledgments

The authors acknowledge many fruitful discussions with Laura Alvarez, Naoufal Elaisati, Lucio Isa, Federico Paratore, Ueli T\"{o}pfer and Carolijn van Baalen for the early parts of this work. Further discussions with Maxime D\'efor\^et Timoth\'ee Gautry, Pierre Illien, Rapha\"el Jeanneret, Christina Kurzthaler, Arnold Mathijssen also helped. S.M. and C. D. are grateful for meeting at the GDR SLAMM.

\bibliography{main_short_v2}

\end{document}